\magnification=1200
\baselineskip=15pt
\hsize=5.5in
{\nopagenumbers
\hskip10cm UM--P-96/98\hfil\par
\hskip10cm RCHEP--96/14 \hfil\par
\hskip10cm Revised Version\hfil\par
\vskip1.5cm
\centerline{\bf Perturbative Prediction for Parton Framentation into
    Heavy Hadron }
\vskip1.5cm
\hskip4cm J.P.  Ma \par
\vskip0.3cm
\hskip4cm School of Physics\par
\hskip4cm Research Center for High Energy Physics \par
\hskip4cm University of Melbourne \par
\hskip4cm Parkville, Victoria 3052\par
\hskip4cm Australia \par
\hskip4cm E-Mail: ma@physics.unimelb.edu.au\par
\vskip2cm
\par\noindent 
{\bf\underbar{Abstract}}:
\par
By expanding functions of parton fragmentation into a heavy hadron 
in the inverse of the heavy quark mass $m_Q$ we attempt to factorize
them into perturbative- and nonperturbative parts. 
In our 
approach the 
nonperturbative parts can be defined as matrix elements
in heavy quark effective theory, the shape of the functions 
is predicted by perturbative QCD. In this work we neglect 
effect at order of $m_Q^{-2}$  and calculate
the perturbative parts at one-loop level for
heavy quark- and gluon fragmentation. We compare our results 
from leading log approximation with experimental results 
from $e^+e^-$ colliders and find a deviation below or at 
$10\%$ level. Adding effect of higher order in $\alpha_s$ 
it can be expected to reduce the deviation. 
The size of matrix elements appearing at the order we consider
for several types of heavy hadrons is determined. 
\par\vskip 20pt
\noindent
\par\noindent
\par\vfil\eject
\ \ \ \ 
\par\vfil\eject}
\baselineskip=15pt
\pageno=1
\noindent 
{\bf 1.  Introduction}
\vskip 15pt
Parton fragmentation functions are in general nonperturbative objects in 
the QCD factorization theorem [1] for predictions of inclusive productions
of single hadron. For a light hadron the fragmentation happens at the energy 
scale of $\Lambda_{QCD}$, which is several hundreds MeV. Hence it is 
purely a long-distance process. For the hadron being a quarkonium, 
which consists mainly of a heavy quark $Q$ and its antiquark $\bar Q$,  
there is in the fragmentation not only long-distance effect but also 
certain short-distance effect because heavy quarks are involved and 
they provide a large energy-scale, the mass $m_Q$ of heavy quarks. This
short-distance effect can be well described by perturbative QCD, while
the long-distance effect can be parameterized with matrix elements 
of local operators defined in nonrelativistic QCD[2]. With these facts 
an amount of functions for parton fragmentation into a quarkonium is 
calculated(See [3-5] for an incomplete list of references). 
A question naturally arises that can we predict  
fragmentation function for a hadron containing a single quark $Q$?  
Here the large mass of the heavy quarks implies certain 
perturbative effect as it does in the case of a quarkonium. 
In this work we attempt to answer the question.  
\par
Recently our understanding of physics related to hadrons containing
single heavy quark has grown rapidly. Such achievement is based on the 
development of an effective theory for heavy quarks, i.e. HQET,  by starting 
from QCD(For HQET see reviews in [6]). 
The basic observation leading to HQET
is that the heavy quark inside a heavy hadron 
carries the most momentum of the hadron. In this work we will refer 
heavy hadrons as those containing single heavy quark. With this observation
one can decompose the momentum of the heavy quark into a large component which 
is roughly the momentum of the heavy hadron, and a small component which 
is at order of $\Lambda_{QCD}$. By integrating out the dynamical
freedom carrying the small component HQET is obtained, in which 
predictions can be expanded in $m^{-1}_Q$, especially, at leading order
every heavy hadron has the same mass---$m_Q$. The difference between 
masses of different heavy hadrons is at order of $m^{-1}_Q$. 
Many applications of HQET have been done for weak decays of heavy hadrons.
Application of HQET to heavy quark fragmentation, i.e., 
heavy hadron production, appeared first in [7], where an expansion 
is obtained, the expansion parameter is the difference 
between the masses of the heavy hadron and of the heavy quark. 
The expansion is nonperturbative and formal, it does not tell
in detail how the fragmentation depends on the energy fraction carried 
by the heavy hadron. 
\par
For heavy quark fragmentation a heavy quark is the initial parton 
which is off-shell with an invariant mass larger than $m_Q$. 
One can image the fragmentation as a two-step process. 
Before combining light quarks and glue 
to form a heavy hadron the heavy quark can emit or absorb light quarks and glue. 
After such emissions it combines light quarks or glue to form 
the hadron and these light quarks and glue should have 
momenta at order of $\Lambda_{QCD}$, as the formation is a 
long-range process. The most momentum of the 
heavy quark at this time will be carried by the hadron 
and the invariant mass of this heavy quark is at the order of
the hadron mass. This heavy qaurk can approximately be treated as on-shell.
Such picture
suggests that the fragmentation function may be written 
in factorized form in which 
a part is for the process from the off-shell- to the on-shell 
quark and another part is for the transition into the 
heavy hadron. The first part can be treated perturbatively 
because of the large scale which is $m_Q$ at least. The second can be parametrized
by using HQET. For other initial partons 
the fragmentation can be thought as a heavy quark is first produced 
and then the formation follows. The production again can be 
handled perturbatively.   
\par
In this work we use the diagram expansion method 
to perform the factorization mentioned above. Such 
method was first used 
in deeply inelastic scattering to analyse twist-4 
effect[8]. This method can be thought as extension 
of Wilson's operator expansion to cases where
the expansion is not applicable, for cases where Wilson's
operator expansion is applicable both methods delivery 
same results.  
For readers unfamiliar with this method 
we refer to [8,9,10]. With the method we
obtain fragmentation
functions as an expansion in $m^{-1}_Q$, in each order
the nonperturbative part is contained in matrix elements 
defined in HQET. In this work we will neglect
all effect which is suppressed by $m_Q^{-2}$. 
For fragmentation function $D_{H/a}(z)$, where 
$a$ stands for the initial parton , $H$ for the heavy hadron 
and $z$ is the momentum fraction carried by $H$, the expansion 
may be written as:
 $$ D_{H/a}(z) = \hat D_a(z) <0\vert O_{H} \vert 0> 
           +O({1\over m_Q^2}).
   \eqno(1.1) $$ 
In Eq.(1.1) $O_{H}$ is an operator 
is defined in HQET, 
its matrix element represents nonperturbative physics. 
The function $\hat D_a(z)$ can be calculated 
perturbatively. We will calculate $\hat D_Q(z)$, $\hat D_G(z)$ 
up to order of $\alpha_s$. The effect at order of $m_Q^{-1}$ is 
included in the matrix element.   
\par
Our work is organized as the following: In Sect.2 we use the diagram expansion 
method for tree-level diagram and HQET to obtain the factorized form
as in Eq.(1.1) for $a=Q$. In doing so, we neglect 
the difference between masses of heavy hadrons and of heavy quarks. 
It should be be noted that at first look one can keep this difference, 
but one  will have some problems at higher order of $\alpha_s$. 
We will explain the problems in detail. In Sect. 3
we proceed to calculate $\hat D_Q(z)$ and $\hat D_G(z)$ 
at order of $\alpha_s$. 
In Sect.4 we 
compare our results with experiment at $e^+e^-$ colliders
and determine the value of several matrix elements $<0\vert O_{H} \vert 0>$
by using $Z$-decays. 
Sect.5 is the summary of our work. 
\par
In this work we will use Feynman gauge and $d$-dimensional 
regularization. In this regularization infrared(I.R.) singularities
appear as poles at $d=4$. We take the normalization of a state 
as: 
  $$ <p'\vert p> =2p^0 (2\pi )^2 \delta ^3({\bf p'} -{\bf p}). 
   \eqno(1.2) $$      
\par\vskip 20pt
\noindent
{\bf 2. The Factorization 
and Tree-Level Results for Heavy Quark Fragmentation }
\par\vskip 15pt
We start in this section our analysis from definitions 
of fragmentation functions. These definitions are first given 
in [11]. Such definitions are conventionally written 
in light-cone coordinate system. In this coordinate system
a $d$-vector $p$ is expressed as $p^{\mu}=(p^+,p^-,{\bf p_T})$, with
$p^+=(p^0+p^{d-1})/\sqrt{2},\ p^-=(p^0-p^{d-1})/\sqrt{2}$.
We introduce a vector  $n$ with
$n^{\mu}=(0,1,0,\cdots, 0)=(0,1,{\bf 0_T})$ in this system. The function of 
heavy quark fragmentation into a heavy hadron $H$ carries momentum $P$ 
is defined 
as[11]: 
 $$\eqalign { D_{H/Q}(z)=& {z^{d-3}\over 4\pi} \int dx^- e^{ -ix^-p^+}
    {1\over 3}{\rm Tr_{color}}{1\over 2}{\rm Tr_{Dirac}}\big [ 
    n\cdot \gamma <0\vert Q(0) \cr
     & \ \ \ \cdot  \bar P {\rm exp}\{-ig_s
  \int_0^\infty d\lambda n\cdot G^T(\lambda n) \}
    a_H^\dagger (P) a_H(P) \cr
  & \ \ \ \cdot 
   P {\rm exp}\{ ig_s \int_{x^-}^\infty d\lambda
        n\cdot G^T(\lambda n)
   \} \bar  Q(x^-n) \vert 0 >\big ] ,  \cr }
    \eqno(2.1) $$
where $G_{\mu}(x)=G^a_{\mu}(x)T^a $, $G_{\mu}^a (x)$ is the
gluon field and $T^a(a=1,\cdots,8)$ are the $SU(3)$-color matrices.
The subscript $T$ denotes the transpose. $Q(x)$ stands for the Dirac-field
of heavy quark. 
$a^\dagger _H(P)$
is the creation operator for the hadron $H$ and $P^\mu=(P^+, P^-,{\bf 0_T})$.  
For hadrons
with nonzero spin the summation over the spin  is understood. The hadron 
carries a fraction $z$ of the momentum $p$ of the heavy quark as the initial parton, i.e., 
$P^+ =z p^+ $. 
The definition is a unrenormalized version. Ultraviolet divergences
will appear in $D_{H/Q}(z)$ and call for renormalization. The renormalization 
is discussed in [11]. We will use modified $MS$-scheme.  
\par 
At tree-level there is only one diagram in the diagram expansion, which 
is given in Fig.1. We divide this diagram with a horizontal broken line
into a upper- and a lower- parts. The upper part contains nonperturbative 
part and we represent it as a black box, corresponding to 
the nonperturbative object $\Gamma_{ij}(q,P)$: 
  $$ \Gamma_{ij}(q,P) =\int d^4 x e^{-iq\cdot x} 
    <0 \vert Q_i(0) a_H^\dagger (P) a_H(P) \bar Q_j(x) \vert 0>, 
  \eqno(2.3) $$
where $i$ and $j$ stand for Dirac- and color-indices. Because of color-symmetry
$\Gamma (q,P)$ is diagonal in color-space. The contribution of Fig.1 
can be written as
  $$ D_{H/Q} (z) = {z \over 24\pi} \int {d^4 q \over (2\pi )^4} 
  {\rm Tr } \big [\gamma\cdot n \Gamma (q,P) \big ] 
   2\pi \delta (n\cdot k- n\cdot q) . 
  \eqno(2.4) $$ 
If we take $H$ to be a quark Q, the black box becomes 
a quark line, and $\Gamma (q,P)$ becomes $(2\pi )^4 \delta^4 (q-P)
(\gamma \cdot q +m_Q)$, we obtain $D_{Q/Q}(z)=\delta (1-z)$. 
The assumption that the hadron carries the 
most momentum of the heavy quark implies that 
the dominant $x$-dependence of the matrix element  
in Eq.(2.3) is $e^{ix\cdot P}$, the correction to this
dependence can be expanded in $\Lambda_{QCD}/m_Q$.    
To proceed further 
we use HQET to write the 
field $Q(x)$ as: 
  $$ Q(x) =e^{-im_Q v\cdot x } \cdot \{ 
   h_v(x) +{1\over 2m_Q}i\gamma \cdot D_T h_v(x)  +O({1\over m^2_Q}) 
  \big \} \eqno(2.5) $$ 
where 
  $$ D_T^\mu = D^\mu -v\cdot D v^\mu, \ \ \  v^\mu = {P^\mu \over M_H} 
 \eqno(2.6) $$
and $m_Q$ is the pole mass of the heavy quark. 
In Eq.(2.6) $D^\mu$ is the covariant derivative. The effective 
Lagrangian for the field $h_v(x)$ reads: 
   $$ L_{\rm eff} = \bar h_v iv\cdot D h_v + {1\over 2m_Q} \bar h_v     
     (i\gamma \cdot D_T )^2 h_v + O({1\over m^2_Q}).  
   \eqno(2.7) $$ 
\par 
The $x$-dependence of $h_v(x)$ in a matrix element is controlled
by the scale at order of $\Lambda_{QCD}$ and can be expanded. 
With Eq.(2.5) and neglecting $M_H-m_Q$ 
the matrix element in Eq.(2.3) can be written as an expansion: 
   $$ \eqalign {<0 \vert Q_i(0) & a_H^\dagger (P)  a_H(0) \bar Q_j(x) \vert 0>
    = e^{+ix \cdot P} \big \{
     <0\vert (h_v)_i(0) a_H^\dagger a_H  (\bar h_v)_j (0)\vert 0>  
     \cr 
   & \  -{1\over 2} x^\mu <0 \vert (\partial_\mu h_v)_i(0) 
         a_H^\dagger a_H  (\bar h_v)_j (0) + (h_v)_i(0)a^\dagger_H a_H 
          (\partial_\mu \bar h_v)_j(0) \vert 0> \cr  
   &\   +{1\over 2m_Q} <0 \vert (i\gamma\cdot D_T h_v)_i(0) 
         a_H^\dagger a_H  (\bar h_v)_j (0)+ (h_v)_i(0) 
          a_H^\dagger a_H \overline {(i\gamma \cdot D_T h_v)}_i(0) \vert 0> 
          \cr & \ + \cdots \big \}. \cr}  \eqno(2.8) $$
The $\cdots$ stand for terms which will lead 
to contributions at order of $m_Q^{-2}$ or higher orders. 
The matrix elements in r.h.s. of Eq.(2.8) are matrices in 
Dirac-indices $i$ and $j$. The structure of these matrices can be 
determined by using symmetries of parity(P), of time-reversal(T). 
The matrix element in the second and third line can be written 
as the form: 
  $$ \eqalign { 
     <0 \vert (\partial^\mu h_v)_i(0)
         a_H^\dagger a_H  (\bar h_v)_j (0) + (h_v)_i(0)a^\dagger_H a_H
          (\partial^\mu \bar h_v)_j(0) \vert 0> &= \cr 
       C_{ij}^\mu  <0 \vert {\rm Tr} \big\{ 
        ((i\gamma\cdot D_T)^2h_v)(0) a_H^\dagger a_H \bar h_v(0) & 
       +{\rm h.c. } \big \} \vert 0>, \cr 
     <0 \vert (i\gamma\cdot D_T h_v)_i(0)
         a_H^\dagger a_H  (\bar h_v)_j (0)+ (h_v)_i(0)
          a_H^\dagger a_H \overline {(i\gamma \cdot D_T h_v)}_i(0) \vert 0> 
       & = \cr 
     B_{ij} <0 \vert {\rm Tr} \big\{  (i\gamma\cdot D_T h_v)(0)
         a_H^\dagger a_H  \bar h_v (0)& + {\rm h.c.} \big \} 
           \vert 0>,  \cr} \eqno(2.9) $$ 
where $C_{ij}^\mu$ and $B_{ij}$ are matrix elements labeled by Dirac-indices. 
Because of $\gamma \cdot v h_v =h_v$ the expectation value in r.h.s.
of the second equation is zero.
In the first equation we used the equation of motion. 
The dimension of the expectation 
value in r.h.s. of the first equation is 3 in $m_Q$, while 
the expectation value in the first line of Eq.(2.8) is 1. 
Hence the contribution from the second line in Eq.(2.8) 
in final results will be at order of $m_Q^{-2}$. 
We define the operator $O_H$ as:  
  $$ O_H= {1\over 12M_H } {\rm Tr} \big \{ 
    h_v(0) a^\dagger _H(P) a_H(P) \bar h_v(0) \big \}. \eqno(2.10) $$
With this operator the nonperturbative object $\Gamma (q,P)$ becomes:  
  $$  \Gamma (q,P)  = (2\pi )^4 \delta^4 (q-P) 
             (\gamma \cdot q +m_Q) <0\vert O_{H} \vert 0> +O({1\over m_Q^2}). 
       \eqno(2.11)$$ 
Finally we obtain
  $$ D_{H/Q}(z) = \delta (1-z) <0\vert O_{H}\vert 0> 
        +O({1\over m_Q^2}) 
       .  \eqno(2.12)$$  
In Eq.(2.12) we have neglected all terms which are at order higher 
than $m_Q^{-1}$. The obtained function is singular at $z=1$. 
If we go to higher orders it becomes more singular, where 
derivatives of $\delta$-function appear. This means 
that the detail of the shape around $z=1$ can not be predicted in our
approach, but physical predictions still can be made by noting
that they are convolutions of fragmentation functions with 
other functions in $z$, the fragmentation functions are 
distributions. With the fragmentation function given in Eq.(2.12) the 
integral with a test function $f(z)$ is approximated by:
  $$ \int ^1_0 dz f(z) D_{H/Q}(z) = f(1)  <0\vert O_{H}\vert 0>
     +O({1\over m_Q^2}) . 
  \eqno(2.13)$$ 
The situation here may look like the case of inclusive 
decays of $B$ mesons[12], 
where one also encounters similar expansion as in Eq.(2.12)
in which terms from higher orders are more singular.
In [12] the accurate shape of  
decay-spectra was interested and the study there is 
corresponding to study the integral in our case: 
  $$ \int ^1 _{1-z_\Lambda} dz f(z) D_{H/Q}(z)   \eqno(2.14)$$ 
for $z_\Lambda$ at order of $\Lambda_{QCD}/m_Q$. Because the order of
$z_\Lambda$ our results in this work can not be applied
to observables which are related to a integral like that in Eq.(2.14).  
The observables studied later are related to the integral in Eq.(2.13).  
\par
One can perform similar analysis as above for other partons. Based
on such analyses we propose to write 
the functions of parton fragmentation into a heavy hadron 
as the factorized form:
 $$ D_{H/a}(z) = \hat D_a(z) <0\vert O_{H} \vert 0>
           +
   O({1\over m^2_Q}).  \eqno(2.15) $$
In this form the hadron dependence is only contained 
in the matrix elements, the $z$-dependence is predicted 
completely by perturbative QCD.  
It should be noted that the matrix element $<0\vert O_{H} \vert 0>$ 
also contains effect at order of $m^{-1}_Q$, because 
it is defined in HQET in Eq.(2.7) with the accuracy at 
order of $m^{-1}_Q$. In our approach here 
the function $\hat D_a(z)$ is just the fragmentation 
function $D_{Q/a}(z)$ of a parton $a$ into a heavy quark $Q$. 
This fact also implies any subtraction for calculating $\hat D_a(z)$
at higher order of $\alpha_s$ 
is not needed. Usually certain subtractions 
are needed to extract perturbative parts in a factorized form. 
To examine this we have calculated the matrix element $<0\vert O_{H} \vert 0>$
by taking $H=Q$ in HQET at one-loop level. Indeed, the matrix 
element does not receive any one-loop correction. This also shows 
that the $\mu$-dependence of the matrix element is suppressed 
at least by $m_Q^{-2}$ or by $\alpha_s^2$. 
\par
In the above approach we have neglect the difference $M_H-m_Q$ in 
the kinematic as in the case with a quarkonium, where 
the binding energy is neglected in the kinematic.  
It seems that such difference can be kept 
in the approach. With the difference HQET given in Eq.(2.6) 
is not suitable for our purpose. The reason is: The heavy quark in Fig.1 
carries the momentum $q=m_Qv +k_1$. In HQET employed above 
we neglect $k_1$ at leading order of $m^{-1}_Q$. The mass $M_H$ 
is usually larger than $m_Q$, so the fragmentation function 
is only non zero at $z=M_H/m_Q$ which is larger than 1. This 
is in conflict with the definition given in Eq.(2.1), which says 
because of the conservation of momenta that
$D_{H/Q}(z) =0$ for $z>1$. This problem may be solved at first loo 
by employing HQET with a residual mass. As already noted 
in [13] that the decomposition of quark momentum $q=m_Qv+k_1$
where $k_1$ is the small component, is arbitrary, one can also 
decompose $q$ as:
  $$ q=(m_Q'+\epsilon_m )v +k_1' \eqno(2.16) $$
where $\epsilon_m$ is the residual mass at order $\Lambda_{QCD}$ 
and $k_1'$ is the small component. With this decomposition  
one can obtain in analogy to Eq.(2.5):
  $$ Q(x)= e^{-i(m_Q'+\epsilon_m)v\cdot x } \cdot \big \{ 
      h_v'(x) + O({1\over m_Q'}) \big \}. \eqno(2.17) $$  
The effective Lagrangian for the field $h_v'$ reads:
    $$ L_{\rm eff} = \bar h_v' (iv\cdot D +\epsilon_m)   h_v' + O({1\over m_Q'}) 
   \eqno(2.18) $$
Repeating the above steps one gets:
    $$ D_{H/Q} = z^3_0 \delta (z-z_0) <0\vert O_{H}'\vert 0> 
         +O({1\over m_Q'^2}),  \ \  z_0 ={M_H \over m_Q'+\epsilon_m}. 
    \eqno(2.19) $$
The operator $O'_{H}$ is obtained by replacing $h_v$ in $O_{H}$ 
with $h_v'$. 
Choosing $\epsilon_m >M_H-m_Q'$ the function is only nonzero 
at $z=z_0<1$.  However, the mass $m_Q'$ is not the pole mass, 
and the pole mass $m_Q$ is the sum $m_Q'+\epsilon_m$. Hence, 
the choice of $\epsilon_m$ is not possible. It seems that 
the effect from the difference $M_H-m_Q$ can not be handled 
in perturbative theory. To study this effect one may only employ
nonperturbative methods or try to sum contributions of a series of
higher-dimensional operators in [7]. 
\par 
In the case of heavy quark distribution in a heavy hadron 
the difference can be kept in 
our approach. The analysis is similar as that leading to Eq.(2.11), 
the corresponding diagram is just this by reversing Fig.1 and $p^+=zP^+$. 
One obtains
     $$ f_{Q/H} (z) =\delta (z-{m_Q\over M_H}) + O({1\over m^2_Q}) 
   \eqno(2.18) $$
where the matrix element corresponding to $<0 \vert O_H \vert 0>$
equals one plus corrections at order of $m^{-2}_Q$. 
The quark line here represents on-shell quark, therefore
the problems mentioned above will not appear.  
\par\vskip20pt
\noindent 
{\bf 3. Results For Fragmentation at One-Loop Level}
\par\vskip20pt
In this section we present a calculation of fragmentation function 
for a heavy quark into a heavy quark and for a gluon into a heavy quark
at order of $\alpha_s$. 
This is the perturbative part in the fragmentation into a heavy hadron 
in Eq.(2.13). In the case of quark fragmentation contribution 
from every diagram at one-loop level contains a I.R. singularity, 
there is a delicate cancellation of the singularity 
between contributions from different diagrams. We show here in detail
how this works.  
\par
At one-loop level there are four diagrams contributing to heavy 
quark fragmentation. They are given in Fig.2A--2D. 
Contribution from each diagram is not only ultraviolet divergent
but also I.R. divergent. However, final result is free from I.R. 
singularity. With the Feynman rule given in [11], the contribution 
from Fig.2A is:
 $$ \eqalign { 
   D_A(z) =& {z^{d-3} \over 24\pi} \mu^\varepsilon 
   \int  ({dk\over 2\pi} )^d {\rm Tr} \big \{ 
    -ig_s T^a \gamma^\mu  
      {i \over \gamma\cdot (q-k) -m_Q+i0^+ } 
     \gamma\cdot n (ig_s T^a n_\mu) 
     \cr  
     &\ \cdot (\gamma\cdot q +m_Q) \big \}  
     \cdot 2\pi \delta (n\cdot (p-q)) \cdot {-i \over 
     k^2 +i0^+}\cdot {i\over n\cdot (p-q+k)+i0^+}  \cr} \eqno(3.1)$$
where $\mu$ is the renormalization scale, $\varepsilon=4-d$, $q^+=zp^+$ and 
$q^2=m_Q^2$.  
The contribution from Fig.2B is:
  $$ \eqalign {
      D_B(z) =& {z^{d-3} \over 24\pi} \mu^\varepsilon
   \int  ({dk\over 2\pi} )^d {\rm Tr} \big \{
    -ig_s T^a \gamma^\mu {i\over \gamma\cdot (q+k)-m_Q+i0^+} 
     \gamma\cdot n (-ig_s T^a n_\mu) \cr
     & \ \cdot (\gamma\cdot q +m_Q)\big \} 
      \cdot {-i\over n\cdot(p-q)-i0^+}\cdot 
   (-1)\cdot 2\pi \delta(k^2) 2\pi\delta (n\cdot (p-q-k)). 
   \cr} \eqno(3.2)$$ 
\par
Both terms are I.R. divergent. In $D_A(z)$ the divergence appears
at $k^+ \sim 0$, while in $D_B(z)$ the divergence is because
the on-shell gluon can carry very small energy. These divergences
can be regularized in dimensional regularization and 
are represented by the terms as $\varepsilon_I^{-1}$, where
$\varepsilon_I=d-4$. Performing the loop integration in Eq.(3.1) and
Eq.(3.2) we obtain: 
  $$ \eqalign { 
      D_A(z) =& -{2\over 3\pi} g_s^2 \delta (1-z) \cdot 
       {\pi^{d-2\over 2} \over (2\pi)^{d-2} } 
      \cdot \Gamma ({\varepsilon \over 2} )     
     ({\mu \over m_Q})^\varepsilon \cr
    & \ \cdot \big\{ {1\over \varepsilon_I } -1 +\varepsilon_I + O(\varepsilon^2)
     \big\} , \cr 
      D_B(z) =& {2z\over 3\pi} g_s^2  \cdot 
       {\pi^{d-2\over 2} \over (2\pi)^{d-2} } 
      \cdot \Gamma ({\varepsilon \over 2} )     
     ({\mu \over m_Q})^\varepsilon \cr
     &\ \cdot \big\{ {1\over \varepsilon_I } \delta (1-z)
           +{1\over (1-z)_+ } 
       +\varepsilon_I \big ( {\ln (1-z) \over 1-z}\big )
   _+ +O(\varepsilon^2) \big\} .\cr}\eqno(3.3)$$   
The $\Gamma$-function $\Gamma ({\varepsilon \over 2})$ with 
$\varepsilon=4-d$ represents 
U.V. divergence. The $+$-prescription is as usual. From Eq.(3.3) 
the sum $D_A(z)+D_B(z)$ is free from the I.R. pole $\varepsilon_I^{-1}$. 
After renormalization the sum is: 
  $$ \big ( D_A(z)+D_B(z)\big)^{(R)} 
       = {2\over 3\pi} \alpha_s(\mu) 
    \big\{ \big[\delta (1-z) +{z\over (1-z)_+} \big ] \ln {\mu^2\over m_Q^2}
     +2\delta(1-z) -2z\big ( {\ln (1-z) \over 1-z}\big )_+ \big \}. 
      \eqno(3.4)$$ 
\par
The contribution from Fig.2C is just the one-loop correction 
to Fig.1. Because the quark-line is for a on-shell quark, the 
correction is to external line. This contribution contains 
also I.R. singularity. After renormalization 
the contribution is:
   $$ D_C(z)^{(R)} = {\alpha_s(\mu) \over 3\pi} 
    \delta(1-z) \cdot \big \{ 
       {2\over \varepsilon_I } +\gamma -\ln (4\pi) 
     -2 +{1\over 2} \ln {m_Q^2 \over \mu^2}  \big \}. 
   \eqno(3.5) $$ 
\par            
The last contribution is from Fig.2D, it reads:  
  $$ \eqalign {      
   D_D(z) &= {z^{d-3} \over 24\pi } \mu^\varepsilon \int 
     \big( {dk\over 2\pi} \big)^d {\rm Tr} \big\{ 
       (-ig_s T^a \gamma^\mu ) {i\over \gamma\cdot (q+k)-m_Q} 
      \gamma\cdot n { -i\over \gamma\cdot(q+k)-m_Q}      
  \cr 
   &\ \cdot (ig_sT^a \gamma_\mu) (\gamma\cdot q+m_Q) \big\} \cdot 
   (-1)(2\pi)\delta(k^2) (2\pi)\delta(n\cdot(p-q-k).\cr } \eqno(3.6)$$ 
Performing the $k$-integration and renormalization we obtain: 
  $$ \eqalign { D_D(z)^{(R)} =& {2\alpha_s(\mu) \over 3\pi} 
         (1-z) \big \{ \ln {\mu^2\over m_Q^2 } -2\ln(1-z) -1 \big\} \cr 
   &\ -{2\alpha_s(\mu) \over 3\pi} \big\{ 
     \big[ {2\over \varepsilon_I}-\ln(4\pi) +\gamma \big]\delta(1-z) 
       +{2z\over (1-z)_+} \big\} .\cr} \eqno(3.7)$$  
The total contribution to the one-loop correction of $D_{Q/Q}$ is
the sum: $2(D_A(z)+D_B(z)+D_C(z))^{(R)}+D_D(z)^{(R)}$. In this sum
the I.R. divergence in Eq.(3.7) cancells that in Eq.(3.5). Therefore
the sum is free from I.R. singularity. With these results the function 
$\hat D_Q(z)$ in Eq.(2.13) is: 
 $$ \eqalign { \hat D_Q(z) =D_{Q/Q}(z) & = \delta(1-z) \cr 
     &\ + {2\alpha_s(\mu)\over 
        3\pi } \big\{ \big[ 1-z +{3\over 2}\delta(1-z) 
               +{2z\over (1-z)_+} \big] \ln {\mu^2\over m_Q^2}
           +2\delta(1-z) \cr 
     &\ +2(1+z)\ln(1-z) +(1+z) -4 \big ( 
       {\ln (1-z) \over 1-z} \big)_+  
           -{2\over (1-z)_+} \big\} .\cr} \eqno(3.8) $$
\par
Now we turn to gluon fragmentation into a heavy hadron. The definition 
of gluon fragmentation function $D_{H/G}(z)$ can also be found 
in [11].  
Upto the order of $m_Q^{-1}$ 
we consider there is only one diagram drawn in Fig.3.  It should be pointed 
out that there are more diagrams at higher orders, in which 
the lower part is connected with the black box not only with the quark  
lines as in Fig.3 but also with some gluon lines. This is also
the case for heavy quark fragmentation if we go beyond the order 
of $m^{-1}_Q$. Repeating 
the procedure in Sect.2 we can obtain 
   $$ D_{H/G} (z)= \hat D_G(z) <0\vert O_H \vert 0> + 
     O({1\over m_Q^2}) \eqno(3.9) $$
as proposed in Eq.(2.13). The function $\hat D_G(z)$ is just 
the fragmentation function $D_{Q/G}(z)$ for a gluon into a heavy quark.  
From Fig.3 the contribution reads  
  $$ \eqalign { & {-z^{d-3} \over 16\pi(d-2)p^+} 
      \mu^\varepsilon \int \big ( {dk\over 2\pi }\big)^d
       2\pi \delta (k^2-m_Q^2) \cr
     &\ \cdot 2\pi \delta(n\cdot(p-q-k)) \big ( {1\over (k+q)^2}\big)^2
      {\rm Tr} \big [ 
        \gamma^\mu (\gamma\cdot k -m_Q)\gamma^\nu (\gamma\cdot q +m_Q) 
       \big ] \cr 
     &\ \cdot (p\cdot n g_{\mu\rho} -n_\mu (k+q)_\rho )
           (p\cdot n g_{\nu\sigma} -n_\nu (k+q)_\sigma) g^{\rho\sigma}. 
    \cr }  \eqno(3.10) $$
We obtain after integration of the loop momentum $k$ and renormalization: 
  $$ \hat D_G(z) =D_{Q/G}(z) = {\alpha_s (\mu ) \over 4\pi} 
      (1-2z+2z^2) \ln { \mu ^2\over m_Q^2 }. \eqno(3.11)$$  
\par
It should be pointed out that the function of heavy quark fragmentation 
into a heavy quark has also been calculated in [14]. In [14] 
the fragmentation function has been extracted by calculating
cross-sections at $e^+e^-$ collider and the result is 
formulated in a compact form
with the $+$-prescription. With the property of the $+$-prescription
one can show that the result in [14] agrees with that in Eq.(3.18). 
\par
For the later purpose we give here some moments for
heavy quark fragmentation into a heavy quark. The moment is defined as
 $$M_Q^{(N)}(\mu) = \int^1_0 dz z^{N-1} D_{Q/Q}(z,\mu). \eqno(3.12)$$ 
The moments for heavy quark fragmentation into a heavy hadron can be obtained 
as $M^{(N)}_{H/Q}(\mu)=
<0 \vert O_H \vert \vert 0> \cdot M_Q^{(N)}(\mu)$ in our approach. 
The first two moments reads
 $$ \eqalign { M_Q^{(1)}(\mu) =& 1 +O(\alpha_s^2 (\mu)), \cr
               M_Q^{(2)}(\mu) =& 1+ {2\alpha_s(\mu) \over 3\pi} 
       \big ( -{4\over 3} \ln {\mu^2\over m_Q^2} -{17\over 9} \big ) 
     +O(\alpha_s^2 (\mu)). \cr} \eqno(3.13)$$ 
These results will be used in the next section. 
\par\vskip20pt
\noindent
{\bf 4. Comparison with experiment at $e^+e^-$ Collider} 
\par\vskip20pt
In this section we will compare our results with experimental results 
obtained from $e^+e^-$ collider. 
Before confronting to experimental results,  we would like to make two 
comments:  
\par
1). If the method for factorization used here works for heavy quark fragmentation, 
it can also be applied directly to single heavy
hadron production without concept of fragmentation. That means that 
one can expand the inclusive cross-section for production 
in term of $m_Q^{-1}$, where the same $\Gamma(q,P)$ in Eq.(2.3) appears
at leading order. In this work we are unable to carry out such program. 
We will still use the results from QCD factorization theorem and neglect 
higher twist effect. In some cases the theoretical analysis 
will be much simple if one uses the results from QCD factorization
theorem and fragmentation functions. 
\par
2). It is confused in the literature about formulation 
in terms of moments of heavy quark fragmentation function for the 
statement that heavy hadron carries the most momentum of the
heavy quark.  
One encounters such formulation for the statement
 $$ M^{(2)}_{H/Q} =\int^1_0 dz zD_{H/Q} (z) = 1 -O({\Lambda_{QCD}\over m_Q}). 
   \eqno(4.1)$$ 
It is easy to see that the formulation is wrong. For the second 
moment one can exactly show that[11]:
 $$ \sum_H M^{(2)}_{H/Q} = 1. \eqno(4.2)$$
With the formulation in Eq.(4.1) the sum-rule in Eq.(4.2) 
can not be hold. Therefore the formulation is wrong. 
The correct formulation for the statement is
 $$ {M^{(2)}_{H/Q} \over M^{(1)}_{H/Q} } =1-O({\Lambda_{QCD}\over m_Q}).   
  \eqno(4.3)$$ 
In our approach the correction terms in Eq.(4.3) begin at order of
$ m_Q^{-2}$, and the factor $1$ receives also radiative corrections
staring at order of $\alpha_s$. 
\par 
If functions for parton fragmentation 
into a heavy hadron is known or extracted from experiment, 
one can calculate the moments of the functions which is usually
at large energy-scale. On the other hand one can calculate the 
moments of the functions using theoretical results as those
in last sections, where one should take the energy-scale $\mu$
to be $m_Q$ to avoid large logarithmic contribution, and then 
use the evolution equation to predict 
the moments at the large energy-scale for comparison 
with the experimental results. Unfortunately, unlike
parton distributions, the functions are not known well 
experimentally. However, for a comparison 
one can calculate directly  
with the theoretical predictions of the functions 
some physical observables, which are well measured in experiment.   
For this purpose we consider the inclusive process
  $$ e^+ +e^- \rightarrow H+ X .     \eqno(4.4)$$ 
Denoting the beam energy as $E_{\rm beam}$ the variable $x_H$ referring
to the hadron $H$ is defined
as: 
  $$ x_H={E_H \over E_{\rm beam}} ={2E_H \over \sqrt{s}} \eqno(4.5)$$ 
where $E_H$ is the energy carried by the hadron $H$. 
With the QCD factorization theorem the differential cross section 
can be written:
 $$ { d\sigma (e^+ +e^- \rightarrow H+ X )\over dx_H } 
     = \sum_a \int ^1_{x_H} {dz\over z} h_a ({x_H\over z}, \mu ) 
       D_{H/a}(z, \mu)  \eqno(4.5)$$ 
where $a$ stands for all possible partons. In Eq.(4.5) 
the function $h_a$ is perturbative part and is known upto one-loop
level in ${\overline{MS}}$-scheme[15]. We will work at one-loop
level and hence we take only the contribution from the 
parton fragmentation where the parton $a$ is the heavy quark $Q$. 
For the fragmentation function we take the results obtained in the 
last section. The function $h_Q$ is[15]: 
  $$ \eqalign{ h_Q(y, \mu) =& 
    \sigma_Q (s) \big\{ 
      \delta (1-y) + {2\alpha_s (\mu)  \over 3\pi} 
       \big\{ \big[  {1+y^2 \over (1-y)_+ } +{3\over 2} \delta (1-y) \
          \big] \ln {s\over \mu^2} \cr
      & \ + ({3\pi^2\over 2} -{9\over 2} )\delta (1-y) 
           +{3\over 2} (1-y) -{3\over 2 } {1\over (1-y)_+} 
        \cr
       &\ +2 {1+y^2 \over 1-y } \ln y 
        +(1+y^2) \big ({\ln (1-y) \over 1-y }\big )_+ +1\big\} \big\}
         \cr}\eqno(4.6)$$  
where $\sigma_Q(s)$ is the total cross-section for 
$e^+e^-\rightarrow Q+\bar Q$ at leading order of coupling constants 
in the standard model. The expectation value of any observable $O(x_H)$
as a function of $x_H$ can be now calculated as
   $$ \eqalign { <O(x_H)> =&  { \int^1_0 dx O(x) 
    \int^1_{x_H} {dz\over z} D_{H/Q}(z,\mu) h_Q({x_H/z}, \mu) \over 
        \int^2_0 dx \int^1_{x_H} {dz\over z} D_{H/Q}(z,\mu) h_Q({x_H/z}, \mu)}\cr
       =& { \int^1_0 dz \int^1_0 dy O(zy)D_{H/Q}(z,\mu ) h_Q(y,\mu) 
       \over \int^1_0 dz D_{H/Q}(z,\mu ) \cdot \int^1_0 dy h_Q(y,\mu)}
       \cr} \eqno(4.7)$$  
If one takes the heavy quark fragmentaion function 
in last sections and neglect the effect at order of $m^{-2}_Q$, 
an interesting consequence is that the measured value of
observable $O(x_H)$ does not depend on the type of heavy hadrons. 
It should be noted that in experiment one can also measure 
$<O(x_Q)>$ by averaging the mean $<O(x_H)>$ for various 
hadrons $H$. The mean $<O(x_H)>$ is predicted by 
replacing $H$ with $Q$ in Eq.(4.7). In our approach we have:
  $$ <O(x_Q)>= <O(x_H)> +O ( {1\over m_Q^2}) 
          =<O(x_{H'})> + O ( {1\over m_Q^2})=\cdots 
      \eqno(4.8) $$
\par
In experiment, the well studied observable is $O(x_H)=x_H$. Recent
measurement at $\sqrt s= M_Z$ by ALEPH[16,17] gives:
  $$ <x_b>=0.715\pm 0.020, \ \ \ \ <x_{H_b}> =0.696\pm 0.016 
       \eqno(4.9) $$
where $H_b$ is the observed hadron in the process 
(4.4) and it can only be $B^0$ or $B^+$. 
These results give certain support 
for Eq.(4.8). A re-analysis of ARGUS data at $\sqrt s=10.6$GeV 
in [18] also shows:
  $$ <x_D> \approx <x_{D^*}> \approx <x_{\Lambda_c}>
         \eqno(4.10)$$ 
at $\sqrt s =10.6$GeV. It is interesting to check whether Eq.(4.8) holds
for othe type of observables in experiment or not. 
In the following we will concentrate
on $<x_Q>$. There are two ways to predict 
$<x_Q>$. One way is only to take the perturbative results in Eq.(4.6) 
and in Eq.(3.13) to calculate $<x_Q>$ in Eq.(4.7). We obtain
  $$ \eqalign { <x_Q>=<x_H> & = {M_Q^{(2)} (\mu) \int_0^1 dy y h_Q(y,\mu) 
            \over M^{(1)}_Q (\mu) \int^1_0 dy h_Q(y,\mu) } \cr 
      & = 1+ {\alpha_s(\mu ) \over 2\pi } \big\{ 
     -{16\over 9} \ln {s\over m_Q^2} +{88\over 27} \big\} + O(\alpha_s^2) 
        .\cr}  
     \eqno(4.11) $$
In Eq.(4.9) there is a large logarithmic contribution. This and those
at higher orders can make the perturbative series unreliable if 
one only take first two or three orders to make numerical
predictions. 
Such logarithmic contributions can however be summed with renormalization 
group equations. In our case we can take $\mu =\sqrt s$ in Eq.(4.7) 
so that the large 
logarithmic $\ln {s\over \mu^2}$ in $h_Q$ disappears. 
For $M_Q^{(N)} (s)$ in Eq.(4.11) we first calculate them 
with the result in the last section at energy-scale $\mu=m_Q$ 
and then use renormalization group equation to obtain $M_Q^{(N)} (s)$.
Here one comment is in order. Since we have already one-loop 
results, one can use renormalization group equations for the 
moments  
at two-loop level to sum not only 
the leading log contributions but also 
next-to-leading log contributions. But the equations 
at two-loop level are unknown. The corresponding 
equations for parton distributions are known at 
two-loop level. At one-loop level there is a simple 
relation between these two sets of equations. At 
two-loop level this relation is not proven to be hold. 
Therefore we take only the renormalization 
group equation at one-loop to sum the leading log contributions 
and tree-level results for $M_Q^{(N)}(m_Q)$, i.e., we take the 
leading log approximation.   
The renormalization group equation for $M_Q^{(N)} (s)$
reads: 
    $$ \mu {d M_Q^{(N)}(\mu ) \over d\mu } 
               = {\alpha_s(\mu) \over 2\pi} 
                     \gamma_{QQ}^{(N)} 
             M_Q^{(N)}(\mu ) + \cdots . \eqno(4.12)$$
In Eq.(4.10) $\gamma_{QQ}^{(N)}$ can be expanded in $\alpha_s(\mu)$ 
and only the leading term is known. 
The leading term for $\gamma_{QQ}^{(1)}$ and $\gamma_{QQ}^{(2)}$
can also be read from Eq.(3.13). We take only the leading term. 
The $\cdots$ stands for the contribution from the moments 
of gluon fragmentation function because of operator mixing. 
Including it the next-to-leading log contributions will be summed. 
In our approach at one-loop level this term should be neglected 
for consistence. 
With these in mind 
we obtain     
  $$ <x_Q> =<x_H> = \left  ( {\alpha_s (m_Q) \over \alpha_s (\sqrt s ) }
   \right  )^{-{32\over 69}} (1  +O(\alpha_s)) \eqno(4.13)$$
where we used one-loop $\beta$-function for $\alpha_s$ and 
5 as flavor number. The terms neglected  
at order of $\alpha_s$ are those terms: a). The terms at order $\alpha_s$
in $h_Q(y, \mu=\sqrt s )$, b). The terms in $M^{(2)}_Q(\mu=m_Q )$
at order $\alpha_s$ and c). The next-to-leading     
term in $\gamma_{QQ}^{(N)}$, two-loop effect in $\beta$-function 
and effect from the moments of the gluon fragmentation function
in Eq.(4.12). To make numerical predictions from Eq.(4.13) 
and Eq.(4.11) we use two-loop $\beta$-function for determining 
$\alpha_s(\mu)$ at different scales. For $\mu\ge m_b$ we take 
5 as flavor number and $\Lambda^{(5)}=200$MeV. With these numbers
we obtain $\alpha_s(M_Z)=0.116$ which is close to the experimental
value measured at $\mu=M_Z$. For $\mu < m_b$ we take 4 as flavor 
number and $\Lambda^{(4)}=400$MeV. With these we obtain 
$\alpha_s(m_\tau)=0.368$ which is also close
the experimental value at $\mu=m_\tau$ where $m_\tau$ is the mass
of $\tau$-lepton. Our input parameters for pole masses of heavy quark 
and for running $\alpha_s$ are: 
 $$ m_b=5.0{\rm GeV}, \ \ \ m_c=1.6{\rm GeV},\ \ \  \Lambda^{(5)}
    =200{\rm MeV},  
\ \ \ \ \Lambda^{(4)}=400{\rm MeV}.  \eqno(4.14)$$
\par
In experiment there are also data for $<x_Q>$ measured 
at $\sqrt s=29$GeV. Several groups have measured $<x_b>$ and $<x_c>$, 
where b- and c- quark were identified with their
inclusive lepton-decays or with charged multiplicity measurements. 
The results from different groups and from different methods
are summarized in [19]. We average these results from 
different groups and from different methods by meaning of unconstrained
averaging as described in [20]. It should be pointed out
that these groups have not only measured $<x_Q>$ but also tried 
to reconstruct the variable $z$ and obtained $<z_Q>$. However, 
such reconstruction relied of Monte-Carlo models for 
fragmentation. We will only make comparison with $<x_Q>$. 
The experimental values which we will compare with our predictions are:
   $$ \eqalign {
          <x_b>=0.715\pm0.020,\ \ \ <x_c>& =0.508\pm 0.011,\ \ \  {\rm for}\  \sqrt s =M_Z, \cr
          <x_b>=0.754\pm0.034 , \ \ \ <x_c>&=0.585\pm 0.036, 
     \ \ \ {\rm for}\  \sqrt s=29{\rm GeV}, \cr
                             <x_c> &=0.640\pm 0.009, \ \ \ {\rm for }\  \sqrt s=10.6
          {\rm GeV}, \cr} \eqno(4.15)$$
where the value for $<x_c>$  
at $\sqrt s=10.6$Gev is from the re-analysis of ARGUS data in [18].
The value for $<x_c>$ at $\sqrt s =M_Z$ is obtained from [21,22] 
where actually the values for $<x_{D^*}>$ are measured, we average them and 
take this value 
as $<x_c>$ according to Eq.(4.8). 
\par
With the input parameters in Eq.(4.14) we obtain from the perturbative 
result in Eq.(4.11) the following numbers: 
 $$ \eqalign { 
          <x_b>=0.867,\ \ \ <x_c>& =0.791,\ \ \  {\rm for}\  \sqrt s =M_Z, \cr
          <x_b>=0.933, \ \ \ <x_c>&=0.843, \ \ \ {\rm for}\  \sqrt s=29{\rm GeV}, \cr
                             <x_c> &=0.906, \ \ \ {\rm for }\  \sqrt s=10.6
          {\rm GeV}, \cr} \eqno(4.16)$$    
where $\mu$ in Eq.(4.11) was taken to be $\sqrt s$. Comparing with experimental
values there are large deviations. The reason is probably  
because large corrections from higher orders in $\alpha_s$ in which
large logarithmic contributions exist. Therefore, the predictions
above may not be reliable.  
With Eq.(4.13) where the leading log contributions are summed 
we obtain:
 $$ \eqalign {
          <x_b>=0.772,\ \ \ <x_c>& =0.563,\ \ \  {\rm for}\  \sqrt s =M_Z, \cr
          <x_b>=0.836, \ \ \ <x_c>&=0.615, \ \ \ {\rm for}\  \sqrt s=29{\rm GeV}, \cr
                             <x_c> &=0.674, \ \ \ {\rm for }\  \sqrt s=10.6
          {\rm GeV}, \cr} \eqno(4.17)$$   
Comparing experimental values for $<x_b>$ the deviation 
from the values given above is $8\%$ at $\sqrt s =M_Z$ and 
$10\%$ at $\sqrt s=29{\rm GeV}$. For the case with c-quark 
the deviation is $10\%$ at $\sqrt s =M_Z$ and is about 
$5\%$ at other energy scales.  
Our predictions here  
are fairly in good agreement with experiment. 
Our predicted values are all larger than experimental values. 
The sources for the deviations mentioned above 
can be various. 
Higher twist effect neglected in Eq.(4.5) and in Eq.(4.7) 
can be one of them.  
An important 
source is the higher order correction in $\alpha_s$. In Eq.(4.13)
we neglected this correction. It should be noted 
that the correction in Eq.(4.13) has the form 
$a\alpha_s(\sqrt s)+b\alpha_s(m_Q)$. Because $\alpha_s(m_Q)$ is
rather large, especially for c-quark, 
this correction can be large. If we add the corrections 
from a). and b). discussed after Eq.(4.13), the deviation at $\sqrt s =M_Z$
is reduced to $5\%$
for c-quark and to $4\%$ for b-quark. At other energy scales 
the reduction is not so significant as that at $\sqrt s=M_Z$, because
this correction becomes smaller as $\sqrt s$ decreases. 
Another possible source is the running $\alpha_s$
at different energy scale, especially, the value of $\alpha_s$ 
at lower energy scales, and also possible nonperturbative 
effect appearing at these scales for running $\alpha_s(\mu)$.  
However, a detail study is needed here. 
\par 
The last question we will study here 
is how large is the matrix element $<0\vert O_H\vert 0>$ defined 
in Eq.(2.9) for a given hadron. This matrix element should be 
calculated with nonperturbative methods, e.g., with lattice QCD. 
It can also be extracted from experimental results. It should 
be noted that this matrix element is universal, i.e., it does not
depend on a specific process. We will use experimental
data obtained in $Z$-decays to extract it for D mesons. However, 
information from experiment is not enough for estimating
these matrix elements uniquely, certain assumptions must
be made. 
For the estimation we do not use the concept of fragmentation. For the inclusive
decay 
  $$ Z\rightarrow H_c +X \eqno(4.18)$$
where $H_c$ stands for $D^0$, $D^{*0}$,
$D^+$ and $D^{*+}$. One can write the 
branching ratio as
  $$ {\rm Br}( Z\rightarrow H_c +X) ={\Gamma_{c\bar c} \over \Gamma_Z} 
     \cdot P(c\rightarrow H_c) + {\Gamma_{b\bar b}\over \Gamma_Z} 
   \cdot P(b\rightarrow H_c) +R_{\rm in} \eqno(4.19) $$
where we neglected the process of the gluon splitting into 
$c\bar c$. The term $R_{\rm in}$ is the contribution
from excited states of $H_c$ which are first produced and then
decay into $H_c$ inclusively.  
With the method here for the factorization the probability
$P(c\rightarrow H_c)$ is just the matrix element:
  $$ P(c\rightarrow H_c) = <0\vert O_{H_c}\vert 0>. \eqno(4.20)$$
The probability $P(b\rightarrow H_c)$ can be expected to be the same 
as $P(c\rightarrow H_c)$ if we consider that the b-quark decays first 
through weak interaction into a c-quark and then the c-quark is transmitted
into the hadron $H_c$. For $D^{*+}$ the ratio of these two 
probabilities is extracted in experiment which is close to 1[21]:
  $$ {P(c\rightarrow D^*) \over P(b\rightarrow D^*)}=1.03\pm 0.21. \eqno(4.21)$$   
We assume that this ratio is one for all $H_c$.  
\par
For the excited state $D^{*}$  we neglect the contribution from $R_{\rm in}$, so we have:
 $$ {\rm Br}( Z\rightarrow D^{*}+X) = {\Gamma_{c\bar c} +\Gamma_{b\bar b}
         \over \Gamma_Z} \cdot <0\vert O_{D^{*}} \vert 0> . \eqno(4.22)$$
There is no information for ${\rm Br}( Z\rightarrow D^{*0}+X)$. 
The matrix element $<0\vert O_{D^{*0}}\vert 0>$ can not be determined 
with Eq.(4.22). If we assume isospin symmetry for light quarks, 
the matrix element is same as $<0\vert O_{D^{*+}}\vert 0>$. 
With the experimental value for ${\rm Br}( Z\rightarrow D^{*+}+X)$
in [20] and isospin symmetry we have:
   $$  <0\vert O_{D^{*0}}\vert 0>=<0\vert O_{D^{*+}}\vert 0>
     \approx 0.22. \eqno(4.23) $$
\par
For $D^+$ we take only the decay $D^{*+}\rightarrow D^+ +X$ 
into account for $R_{\rm in}$, according to [20] the decay
has a chance of $31.7\%$. With that we have
  $$ {\rm Br}( Z\rightarrow D^{+}+X)={\Gamma_{c\bar c} +\Gamma_{b\bar b}
         \over \Gamma_Z} \cdot \big \{
      <0\vert O_{D^{+}} \vert 0> +31.7\% <0\vert O_{D^{*+}}\vert 0> 
     \big \} . \eqno(4.24)$$
Taking experimental value we obtain
  $$ <0\vert O_{D^{+}} \vert 0> \approx 0.39 .\eqno(4.25)$$
For $D^0$ we take only the contribution for $R_{\rm in}$ 
from the two decays $D^{*0},\ D^{*+} \rightarrow D^0 +X$. The branching
ratio for these two decays is $100\%$ and $68.3\%$ respectively[20]. With these
the branching ration ${\rm Br}(Z\rightarrow D^0 +X)$ can be written:
  $$ {\rm Br} (Z\rightarrow D^0 +X) = {\Gamma_{c\bar c} +\Gamma_{b\bar b}
         \over \Gamma_Z} \cdot \big \{ <0\vert O_{D^0} \vert 0> 
         +<0\vert O_{D^{*0}}\vert 0> +68.3\% <0\vert O_{D^{*+}}\vert 0> 
    \big\}. \eqno(4.26)$$
It is interesting to note that with isospin symmetry as assumed      
before the matrix element with $D^0$ is the same as that in Eq.(4.25)
and this branching ratio is predictable with Eq.(4.26).    
We obtain 
   $$ {\rm Br} (Z\rightarrow D^0 +X) \approx 20.1\% \eqno(4.27)$$ 
which is close to the experimental value $20.7\pm 2.0\%$[20], or using the
experimental value we obtain
   $$ <0\vert O_{D^0} \vert 0> \approx 0.40 \eqno(4.28)$$  
which is in consistence with the assumed isospin symmetry. 
\par
Naively one would expect the matrix element $<0\vert O_D \vert 0>$
to be 3 times of $<0\vert O_{D^*}\vert 0>$ because $D$ and $D^*$ are spin-0 
and spin-1 particles respectively. From our estimation above this relation
is not hold. The reason for this is that the spin counting 
can not be applied here because the matrix element $<0\vert O_{H_c}\vert 0>$
is the probability for the inclusive transition $c\rightarrow H_c +X$, 
where the unobserved state $X$ can not be vacuum or a given state 
and can have any possible orbital 
angular momentum. Further, the unobserved state $X$ for $D$ can be different 
than that for $D^*$.  
\par   
There is not data available for b-flavored hadrons, so their matrix
elements can not be estimated as we did for $<0\vert O_{H_c}\vert 0>$. 
However the difference between $<0\vert O_{H_c}\vert 0>$ and  
$<0\vert O_{H_b}\vert 0>$, where $H_b$ stands for $B$ or $B^*$ mesons,  
is at order of $m_c^{-1}$ and of $m_b^{-1}$. Therefore
one can take the value of $<0\vert O_{H_c}\vert 0>$ for the corresponding
$<0\vert O_{H_b}\vert 0>$ as an approximation. 
\par\vskip 20pt
\noindent
{\bf 5. Summary}
\par\vskip 20pt
In this work we studied parton fragmentation into a heavy hadron. 
We factorized the process into a perturbative part and a nonperturbative
part. The perturbative part is just the parton fragmentation function 
into a heavy quark at the order we consider. The nonperturbative
part is a matrix element defined in HQET, which is universal. 
The $z$-dependence of fragmentation functions 
is predicted purely by perturbative theory.  
In this work we predicted this dependence for heavy quark- and gluon-  
fragmentation at one-loop level in QCD. With these results
we calculated the mean value of the ratio between the energy carried by 
a heavy hadron or a heavy quark and the beam energy at $e^+e^-$ colliders. 
Comparing experiment we find that there is a deviation at $10\%$ level 
between our predictions with the
leading log approximation  and experimental values. The sources for
this deviation can be several, the important source may 
be higher order effect in $\alpha_s$ as discussed in the last section. 
The value of the matrix element for $D$ and $D^*$ are  
determined with experimental data from $Z$-decays and 
one of the branching ratios for $Z\rightarrow H_c +X$ 
can be predicted in our approach. 
\par
It should be pointed out that our procedure for factorization 
may directly be applied to heavy hadron production with the concept
of fragmentation. In this work we do not carry out
such program for specific process and leave it for future work.  
\par\vskip20pt
\noindent
{\bf Note added}:
\par
After the work is finished, the author is informed by Prof. O. Biebel 
of OPAL group about recent measurement of the total branching 
of $c\rightarrow D^*$. The latest preliminary 
value of the measured branching $f(c\rightarrow D^* +X)$ 
is $0.221\pm 0.014 \pm 0.013$[23]. In the approach of our work 
the branching $f(c\rightarrow D^* +X)$ is just the matrix element
$<0\vert O_{D^*}\vert 0>$. The value obtained in this work 
in Eq.(4.23) is close to the value measured by OPAL.   
\par\vskip 20pt
\noindent    
{\bf Acknowledgment:}
\par
The author would like to thank Prof. W. Bernreuther and Prof. O. Biebel 
for communications. This work is supported by Australian Research 
Council. 

\par\vfil\eject
\centerline {\bf Reference}
\par\noindent 
[1] J.C. Collins, D.E. Soper and G. Sterman, in Perturbative Quantum
Chromodynamics,
\par\noindent \ \ \ \ edited by A.H. M\"ulller, World Scientific, Singapore,
1989.
\par\noindent
[2] G.T. Bodwin, E. Braaten and G.P. Lepage, Phys. Rev. D51 (1995) 1125
\par\noindent
[3] E. Braaten and  T.C. Yuan, Phys. Rev. Lett. 71 (1993) 1673
\par\noindent
\ \ \ \ E. Braaten and T.C. Yuan, Phys. Rev. D50 (1994) 3176
\par\noindent
\ \ \ \ E. Braaten, K. Cheung and T.C. Yuan, Phys. Rev. D48 (1993) 5049
\par\noindent
[4] A.F. Falk, M. Luke, M.J. Savage  and M.B. Wise, Phys. Lett. B312 (1993)
486 
\par\noindent
\ \ \ \ P. Cho and M.B. Wise, Phys. Rev. D51 (1995) 3352
\par\noindent
\ \ \ \ P. Cho, M.B. Wise and S.P. Trivedi, Phys. Rev. D51 (1995) 2039 
\par\noindent
[5] J.P. Ma, Phys. Lett. B332 (1994) 398, Nucl. Phys. B447 (1995) 405,  
\par\noindent 
\ \ \ \  Phys. Rev. D53 (1996) 1185 
\par\noindent
[6] M. Neubert, Phys. Rep. C245 (1994) 259 
\par\noindent
\ \ \ \ M. Shifman, Lectures given at Theoretical Advanced 
Study Institute QCD and Beyond, 
\par\noindent 
\ \ \ \  University Colorado, June 1995, hep-ph/9510377
\par\noindent
[7] R.L. Jaffe and L. Randall, Nucl. Phys. B412 (1994) 79 
\par\noindent
[8] R.K. Ellis, W. Furmanski and P. Petronzio, Nucl. Phys. B212 (1983) 29
\par\noindent
[9] H.D. Politzer, Nucl. Phys. B172 (1980) 349
\par\noindent
[10] J. Qiu, Phys. Rev. D42 (1990) 30 
\par\noindent   
\ \ \ \  \ J. Qiu and G. Sterman, Nucl. Phys. B353 (1991) 105, {\it ibid} 137 
\par\noindent
[11] J.C. Collins and D.E. Soper, Nucl. Phys. B194 (1982) 445 
\par\noindent
[12] M. Neubert, Phys. Rev. D49 (1994) 4623
\par\noindent
\ \ \ \ \ I. Bigi, M. Shifman, N. Uraltsev and A. Vainshtein, 
  Int. J. Mod. Phys. A9 (1994) 2467 
\par\noindent
[13] A.F. Falk and M. Neubert, Nucl. Phys. B388 (1992) 363 
\par\noindent 
\ \ \ \ \  F.E. Close and Zhenping Li, Phys. Lett. B289 (1992) 143
\par\noindent
[14] B. Mele and P. Nason Nucl. Phys. B361 (1991) 626, Phys. Lett. B245 (1990) 635 
\par\noindent
[15] G. Altarelli, R.K. Ellis, G. Martinelli and S.Y. Pi, Nucl. Phys. B160 
(1979) 301
\par\noindent
\ \ \ \ \ R. Baier, K. Fey, Z. Phys. C2 (1979) 339 
\par\noindent
[16] D. Buskulic et al, ALEPH Collaboration, Phys. Lett. B357 (1995) 699
\par\noindent
[17] G. Alexander et al, ALEPH Collaboration, Phys. Lett. B364 
(1995) 93 
\par\noindent
[18] L. Randall and N. Nius, Nucl. Phys. B441 (1995) 167
\par\noindent 
[19] J. Chrin, Z. Phys. C36 (1987) 163
\par\noindent
[20] R.M. Barnett et al, Particle Data Book, Phys. Rev. D54 (1996) 
\par\noindent
[21] R. Akers et al., OPAL Collaboration, Z.Phys. C67 (1995) 27
\par\noindent
[22] D. Buskulic et al., ALEPH Collaboration. Z. Phys. C62 (1994) 1
\par\noindent 
[23] OPAL Physics Note PN 227, preliminary results of OPAL submitted 
to the ICHEP
\par\noindent 
\ \ \ \ \ Conference, Warsaw, 
July,1996, and the DPF Conference, Minneapolis, August.     
\par\noindent
\par\noindent
\vfil\eject
\centerline{\bf Figure Caption}
\par\vskip20pt
\noindent
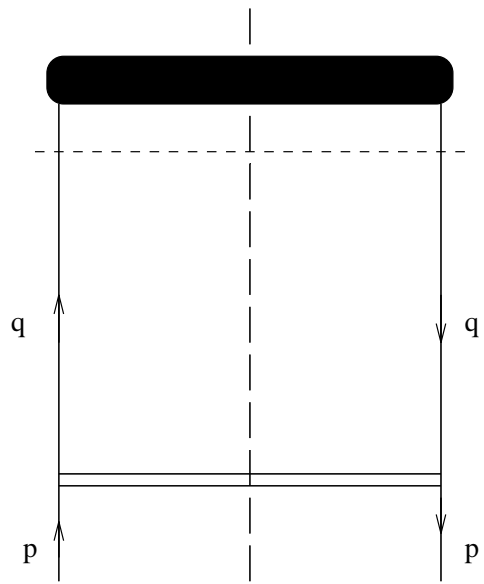
{\bf Fig.1}: The Feynman diagram for contribution to heavy quark fragmentation 
at tree-level. The vertical broken line is the Cutkosky cut, the 
double line presents the line operator in Eq.(2.1). 
\par\noindent
{\bf Fig.2A--2D}: The Feynman diagrams for one-loop contributions 
to heavy quark fragmentation into a heavy quark. 
\par\noindent
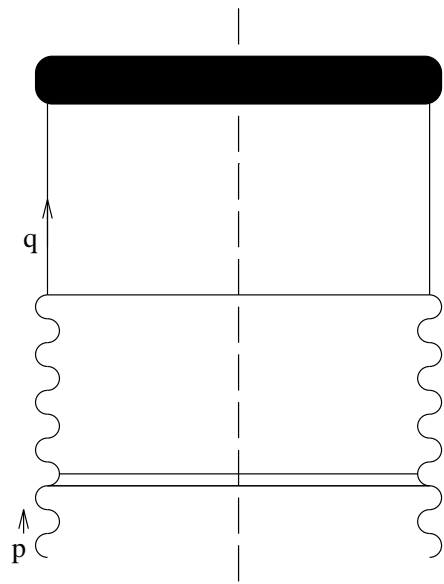
{\bf Fig.3}: The Feynman diagram for contribution to
gluon fragmentation into a heavy hadron. 
\par\vfil\eject
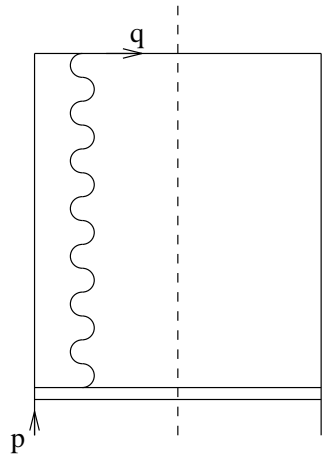
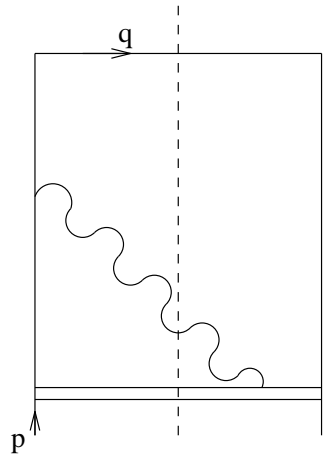
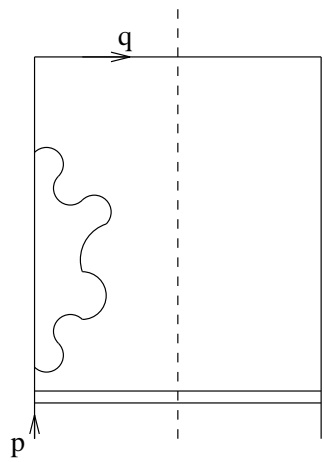
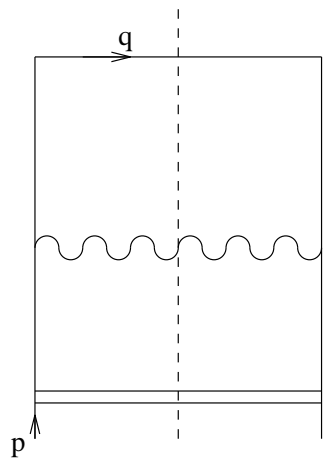
\end